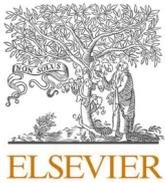

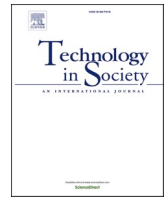

# A comparison of primary stakeholders' views on the deployment of biometric technologies in border management: Case study of SMart mobILity at the European land borders

Mohamed Abomhara [a, *], Sule Yildirim Yayilgan [a], Livinus Obiora Nweke [a], Zoltán Székely [b]

[a] Department of Information Security and Communication Technology, Norwegian University of Science and Technology (NTNU), Gjøvik, Norway
[b] Szekely Family and Company Kft, Budapest, Hungary

### ARTICLE INFO

### ABSTRACT

Advances in technology have a substantial impact on every aspect of our lives, ranging from the way we communicate to the way we travel. The Smart mobility at the European land borders (SMILE) project is geared towards the deployment of biometric technologies to optimize and monitor the flow of people at land borders. However, despite the anticipated benefits of deploying biometric technologies in border control, there are still divergent views on the use of such technologies by two primary stakeholders–travelers and border authorities. In this paper, we provide a comparison of travelers' and border authorities' views on the deployment of biometric technologies in border management. The overall goal of this study is to enable us to understand the concerns of travelers and border guards in order to facilitate the acceptance of biometric technologies for a secure and more convenient border crossing. Our method of inquiry consisted of in-person interviews with border guards (SMILE project's end users), observation and field visits (to the Hungarian-Romanian and Bulgarian-Romanian borders) and questionnaires for both travelers and border guards. As a result of our investigation, two conflicting trends emerged. On one hand, border guards argued that biometric technologies had the potential to be a very effective tool that would enhance security levels and make traveler identification and authentication procedures easy, fast and convenient. On the other hand, travelers were more concerned about the technologies representing a threat to fundamental rights, personal privacy and data protection.

## 1. Introduction

Over the past few decades, there has been a significant increase in the number of travelers crossing borders. The World Tourism Organization (UNWTO) has stated in its long-term forecast report that the number of people crossing international borders is expected to reach 1.8 billion by 2030 [30]. This influx of people is already straining the resources at borders around the world. The situation is further exacerbated by security concerns over different threats from terrorism through to human trafficking and the pandemic, which has led various governments to seek better methods for securing borders [10]. For instance, on April 6, 2016, the European Union (EU) Commission introduced an improved package (the "Smart Borders" package) to establish an Entry/Exit System (EES) for border control [8,19]. The aim of the Smart Borders package is to enhance EU border management as well as improve the effectiveness and efficiency of border control through the adoption of new technologies.

These technologies include large-scale biometric information systems combined with personal identity verification systems [26]. Examples of these systems include the Visa Information System (VIS), the Second-generation Schengen Information System (SIS II), and the European Asylum Dactyloscopy Database (EURODAC) [21]. Most of these systems employ automated, multimodal, biometrics-based identification and verification for personal recognition. Hence, there have been massive deployments of such systems in Automated Border Control (ABC) [14,20], typically through so-called e-gates, in many EU ports of entry in recent years [15].

Accordingly, the SMart mobILity at the European land borders (SMILE) project is geared towards achieving interoperability with other border information systems. SMILE is a promising approach in terms of enhancing the speed, efficiency and flow of border crossing as well as border security. SMILE, like other EU information systems, has the

---

tendency to collect, use and process personal data (e.g., alphanumeric data, such as names and birth dates, and biometric data such as fingerprints) in order to optimize and monitor the flow of people at land borders. Its services have significant potential to make border crossing more effective and efficient as well as to enhance security and detect fraud.

Despite the potential benefits of using biometric technologies in border control, inappropriate use of such a technology may corrode public trust due to ethical, privacy and data protection challenges [2,4, 13,17,23,24,29]. Biometric technologies tend to generate more information than is needed for a specific purpose, and create the danger of using this additional information for other purposes for which it has not been intended or authorized [18]. This type of situation poses serious concerns regarding the fundamental rights such as right to privacy and data protection of travelers crossing the borders [17,29].

Furthermore, the growing proliferation of the use of biometric technologies in border control cannot be justified by their potential benefits alone. A balanced view of important stakeholders (travelers and border authorities) is required to facilitate the acceptance of such technologies. If a majority of the travelers (who are, in fact, clients of border control authorities) do not consider the compulsory use of biometrics proportional to the benefits gained by the improvements they afford, then the arguments for often costly deployment of these technologies are significantly undermined.

In this paper, we examine the views of travelers and border authorities on the deployment of biometric technologies in border management. Our qualitative method of inquiry includes interviews, field observation of two EU land borders and questionnaires. The overall goal of this study is to enable us to understand the concerns of the travelers and border authorities with respect to the use of biometric technologies in border management, and to recommend best practice approaches that may be employed to address concerns such as right to individual privacy and data protection. This would facilitate the acceptance of biometric technologies for a secure and more convenient border crossing.

The remaining part of this paper is organized as follows. Section 2 presents the use of biometric technologies in border management and their potential benefits. Section 3 describes the methods employed to obtain the views of travelers and border authorities on the deployment of biometric technologies in border management. Section 4 presents the results of the inquiry, followed by a discussion of the outcomes in Section 5. Section 6 concludes the paper.

## 2. Background

Organizations responsible for border security have always been under pressure to follow the development of technology in order to maintain their ability to control the border. Biometric technologies of the present are the next stage in the long history of travel documents originally connected with trade development. Identification started with having a host person responsible for one's deeds towards the local authority (called *proxeny* in ancient Greece), followed by clay tokens whose holder was allowed to enter. The first written document recognized as a passport was issued in 450 BC, and afforded its bearer free passage. The verbiage used in such ancient documents is still used in modern passports given its diplomacy. Over the centuries, abuse and misuse of passports (including unintended) spread, and interest in strict control of borders rose. Passports started to provide increasingly accurate information on their holders: name, nationality, height, gender, hair color, skin color, eyes, religion, and spoken languages. In the 19th century, emerging nationalism and absolutism culminated in the institution of "police state", which brought about a previously unprecedented stringency in border control as state borders became increasingly important symbols of sovereign national states. This control quickly adopted state-of-the-art technologies. In less than 20 years after photographic identification was first demonstrated during the 1879 Centennial Exposition in Philadelphia, the first passports containing photographic images appeared, and by the end of the Great War it became a de facto standard. The development of mobility boosted the volume of business and leisure travel, and facilitating border crossing came to be in countries' best economic interests, counterbalancing the national security interest in strict border control. The introduced and currently existing technology of mass border check procedures requires a lot of human resources, equipment and infrastructure from the implementing countries. This is even true in the present, because although the current pandemic has decreased travel volumes, frequency of controls, for example temporary border checks at Schengen internal borders, are stretching existing capabilities.

Generally speaking, a shift of balance towards automation perceived simply just as a normal course of development in any branch of industry or service market. Use of biometric technologies is a key driver in this shift, as it facilitates identification of humans by machines, enabling the new technology of Automated Border Control (ABC) [14,20]. This enables border authorities to regroup human resources from mass border checks (called a "first line" check) to special cases ("second line" or "thorough" checks) where more attendance is needed (for example to detect victims of child trafficking at the border).

Similar to deployment of the Internet of Things (IoT) for security in other fields, the introduction of biometric identification technologies can result in a shift of threats and risks: while it decreases threats to integrity and human-borne risks (e.g., infection, corruption, negative discrimination, mistreatment, human error, and disproportionate use of force), it increases vulnerability to cyber threats and the risk of loss of privacy through data incidents or abuse by data controllers or data processors. A recent example of this is a May .31st. breach on US Customs and Border Protection that exposed the biometric datain this case facial images of "less than 100,000" people.

In sum, border control authorities, as stakeholders in the deployment of biometric technologies in border management, have a great interest in the exploitation of automated border checks facilitated by biometric technologies. They must, however, take serious responsibility for counterbalancing the threats and risks resulting from such a shift.

### 2.1. The use of biometric technologies in border management

There has been a growing deployment of biometric technologies across the EU borders, which can be attributed to increasing numbers of international travelers. In fact, border crossings in and out of the EU is expected to reach 887 million by 2025 [8]. However, border control resources can no longer keep pace with the increasing number of travelers crossing This has led to the introduction of Smart Borders policy by the European Commission in 2011 [8,19].

The Smart Borders policy aims to employ new technologies like automated information sharing and self-service for a smoothly-running of border crossings and securing the EU's external borders [15,19]. It achieved this through an April 2016 legislative proposal for a common entry/exit system (EES) for third country nationals [8]. Also, the Passenger Name Record Directive [16] was proposed to enhance security features in travel documents, more systematic checks at the EU external borders and improvement of the Schengen Information System. These legislative proposals paved the way for the use of large-scale biometric information systems combined with personal identity verification systems for border control.

ABC [14,20] utilizes biometric technologies (which exploit physiological and behavioral characteristics for identification and verification) as a means of identifying and verifying an individual's identity. The method used at the ABC involves the verification of an individual's identity by comparing the new template generated at the automated biometric gate (e-gate) to the enrolled template from the biometric samples stored in an electronic document such as the e-Passport [3]. This is the opposite of the manual border checks where the identification of travelers and the verification of their travel documents are done by border guards using technical equipment like document scanners.

However, methods automated or manual are based on Schengen Borders Code (SBC) regulation [9] and must follow a strict procedure. Automated border control differs from manual control in three ways: First, in the automated check, the passenger puts their own document on the scanner. Second, identification and verification is done by a software. Third, if there is any doubt, the ABC e-gate does not ask questions before sending the passenger to the second line check. Thus, automated border control has been touted to have the potential for speeding up the border crossing process to ensure a secure and more convenient border crossing.

Although the use of biometric technologies has the potential to improve the border crossing process, it still raises some ethical, legal and social implications that need to be considered [4,5,13,17,23–25,29]. Biometric technologies have been shown to generate more interest in border than required for a specific purpose, and there is the danger that such additional information may be exploited for unintended or unauthorized purposes [18]. Also, it may facilitate discriminatory social profiling [28] and can be invasive [31]. Other issues that the use of biometric technologies in border control could raise involve human dignity and children's rights [1].

Indeed, the use of biometric technologies in border control has come to stay. Governments around the world are currently implementing different variants of biometric technologies to support the ever increasing number of travelers crossing the borders and limited border control resources. The case study of the EU's deployment of biometric technologies in border control, presented in this subsection, is an indication that more governments around the globe are likely to follow suit. However, it is important to ask why the use of biometric technologies is generating such hype? What exactly are the potential benefits of using biometric technologies in border control? To respond to these questions, we present the potential benefits of biometric technologies in border control in the next subsection.

### 2.2. Potential benefits of deploying biometric technologies in border checks

There are several potential benefits that could be derived from the use of biometric technologies in border checks. Both travelers and border authorities desire a secure and more convenient border crossing experience. However, the goals of these two groups represent two sides of the same coin. On one side of the coin, travelers are interested in a stress-free border crossing experience. Meanwhile, on the other side of the coin, border authorities are interested in using whatever means necessary to ensure border security. The attainment of these two goals implies that trade-offs would have to be made in terms of the methods to apply. Biometric technologies have the potential to meet both objectives. This can be done through ensuring the accuracy, integrity, robustness, and efficiency of the overall border crossing process, which would therefore lead to a more convenient border crossing experience (meeting the needs of travelers) and a secure border checks (addressing the concerns of border authorities).

The accuracy of the border crossing process refers to the ability to correctly recognize the identity of travelers and reject impersonators. As of now, the processes of identification and verification are done manually by border guards. It relies on human perception to acquire knowledge about the traveler and associate such knowledge with the identity of the traveler as represented by their travel documents (e.g., passport). Unfortunately, the accuracy of the current means of identification may be affected by several factors such as lighting, age of the picture, perception capabilities of the border guard, fatigue, etc. [1]. However, the use of biometric technologies has the potential to enhance and support these processes. For example, it has been shown that the use of multimodal biometric identification resulted in a higher average accuracy throughout the duration of usage, and it also facilitated cross-checking of personal data with greater accuracy [12].

Another potential benefit of using biometric technologies in border control is its ability to ensure integrity in the process. Integrity as it relates to border control has to do with the ability to ascertain whether the presented document and its content have been altered in any way [12]. The factors that affect the accuracy of border guards are likely to affect their ability to verify the integrity of the presented travel document and its content. Hence, the use of biometric technologies in border control can reduce the likelihood of a traveler with an altered travel document being allowed across the border. This is because biometric verifications do not depend on the border guards, but rather on automated verification mechanisms. However, travel documents that are forged or obtained through bribery used by the most dangerous players are undetectable by current ABC solutions in Europe. A recently finished EU project called iBorderCtrl[1] as well as the AVATAR system in the US are already addressing this issue with different deception detection technologies.

Furthermore, biometric technologies are robust in the sense that they are easy to operate, maintain, update, replace, redeploy or decommission when compared to border control units/booths requiring human agents [1]. Ordinarily, for border control with border guards, it takes a long time for agents to become competent, usually through repetitive training. In the case of border control with biometric technologies, such timeframes for achieving competency are not required. In addition, the use of biometric technologies could offer travelers alternatives in identification and authentication methods. For instance, a traveler could decide (theoretically) which biometric modality (e.g., fingerprints, face, iris) they wish to use for identification [1].

The efficiency of deploying biometric technologies in border control can be seen in the ability of automated border control gates to maintain their processing capability throughout their period of operation. This is unlike border controls operated by border guards, as it has been observed that while a border guard is usually efficient at the start of their shift, their efficiency tends to decrease over time [12]. It can then be concluded that using biometric technologies in border control could offer much more accurate and efficient border checks when compared to border controls without such technology.

### 2.3. Legal framework and requirements

During the requirements analysis phase, the SMILE research project took into consideration EU (e.g., General Data Protection Regulation (GDPR) [11,27]) and national (Hungarian, Romanian and Bulgarian) laws on border control, as well as laws on privacy and data protection. To simplify a systemic review of EU and national laws, questions related to the legal (privacy and data protection) requirements applicable to the processing of personal data by border authorities were submitted to border authorities' (Hungarian, Romanian and Bulgarian) legal departments. This yielded a comprehensive description and analysis of the legal and policy requirements regarding travelers' personal consent; hosting/storage of personal data; creation, access to, and update of personal data; accountability; secondary uses and archiving duration of data; and legal requirements on interoperability with other EU information systems. This helped both the SMILE team and border authorities to describe legal measures needed to achieve best practices in data protection, as well as comply with the requirements for EU and national data protection.

A preliminary finding of the SMILE legal review showed that although biometric data processing is the subject of extensive legislation in the EU, analysis of the current European legal framework shows that the legal basis for biometrics processing at the border is not in all cases explicit. The answers obtained from the systemic review demonstrated that there is no single general legal framework governing the processing of biometric data at the border. While processing personal data, border authorities are required to comply with multiple regulations such as the

---

[1] https://www.iborderctrl.eu/.

Schengen Border Code (SBC) and other regulations that govern border control systems such as SIS II regulations and VIS regulations, among others. In addition, the GDPR and Directive 2016/680 [22] for data protection encourage the development of biometric technologies to understand that data privacy and protection are not just a matter of simple compliance. Thus, border authorities' policy for biometric data collection and processing must be based on legal principles and involve the participation of diverse actors. Also, biometric technology deployments such as SMILE must involve control mechanisms that help ensure that all rights of data subjects are respected.

## 3. Methods of inquiry

In this section, we present methods employed to obtain the views of travelers and border authorities on the deployment of biometric technologies in border management. The methods consist of following:

- **Field observation and interview:** The SMILE team visited two different land Border Crossing Points (BCPs) to directly observe the border control processes in order to understand the nature of tasks and the context in which the border guards perform their duties. Also, during these two visits, the border guards were interviewed by the SMILE team to discuss their thoughts and gain insight into their perspectives on business needs and the feasibility of potential solutions to facilitate a secure and more convenient border crossing.
- **Questionnaires:** SMILE conducted two web-based questionnaires: one for land border guards and one for travelers, to investigate and understand participants' views on the use of biometric technologies used in the border's management.

### 3.1. Field observation and interview

On March 8, 2018, the SMILE team visited the Hungarian -Romanian BCP (Nădlac) and observed the workflow of the border crossing. Nădlac (called *Nagylak* in Hungarian) is the main BCP into western Romania from Hungary. Moreover, on April 12, 2018, the SMILE team visited the Bulgarian-Romanian BCP in Ruse. The Bulgarian-Romanian border stretches over 609 km, 470 km of which is formed by the Danube River. Along this border there are 14 BCPs: seven by boat, five by road and 2 by rail.

The observers from the SMILE team were inside and outside of the BCPs' booths observing the regular check performed by the border guards, taking notes and asking questions when clarification was needed. Also, the interview process was designed to identify the needs of border guards so that the complication of SMILE system requirements would be as close to the end-user needs as possible. The SMILE team conducted an unstructured interview (one-on-one interviews and group interviews) with land border guards at Nădlac and Rusa BCPs. In the interview sessions, the SMILE team attempted to achieve a holistic understanding of the workflow, develop a real sense of a person's understanding of a situation as well as uncover a rich set of requirements in a short period of time. An unstructured interview was chosen because the SMILE team wanted the interviewees (duty officer, chief of BCP, etc.) to feel free to talk about what they deemed important with additional information requested by the interviewer if it was needed.

### 3.2. SMILE questionnaires for border guards

The SMILE questionnaires targeted border guards such as passport control officers, officers in duty and heads of shift, to name a few. The sample of respondents was a convenience sample [6,7], where our target population met certain practical criteria, such as easy accessibility, geographical location, and/or the willingness to participate in the survey.

The goal was to obtain the views of different border guards at the land borders on the use of biometric technologies in border crossing. As a result, there were 40 respondents to the SMILE questionnaires for border guards. The respondents distribution was based on their position on the land border. Fifty percent of all respondents (20 individuals) were passport control officers, while 10% (4 individuals) were document/ vehicle experts. Three individuals were heads of department, three were duty officers, three were heads of shift, and two were senior passport control officers. The remaining five individuals fulfilled roles as crossing point commanders, chief of communications, or informatics officers. Moreover, 50% of respondents had more than 10 years of experience (between 15 and 25 years of experience); 37.5% had 5–10 years of experience, and 7.5% had between 1 and 5 years of experience.

The questionnaires were coded into three categorical variables to help understand and examine the viewpoint of border guards with respect to the use of biometrics and its benefits in border control. The categorical variables were personal data used for identification and verification; information sharing, storage and legislation; and visa and passport handling during crossing/entering borders.

### 3.3. SMILE questionnaire for public travelers

There were 90 respondents to the SMILE questionnaires for travelers. The gender distribution of the respondents was unfortunately uneven, as can be observed in (Fig. 1), where the overwhelming majority of them (74.4%) was male.

Respondents ranged in age from 18 to 64 years (Fig. 2). The biggest respondent group was in the 35–44 age range (34.4%). Following this group, we observed that the second largest group was slightly younger, and fell into the 25–34 age range (30.0%). Overall, the majority of travelers consisted of people under 45 years of age (80%).

The participants were asked questions related to their travel habits to countries within the European Union and any problems they had experienced during their trips. The first question they were asked was "how often do you travel through the EU land border?" The distribution of respondents by frequency of traveling through EU land border revealed that 42.2% of respondents traveled a few times a year (less than 3 trips per year). The second largest group (25.6%) reported moderate annual travel; 17.8% claimed to travel very frequently; and only 14.4% reported that they had never traveled through the EU land border.

## 4. Results

### 4.1. Field observation and interview

During the land border visits and interview, a wide range of information was gathered by the SMILE team. According to the borders' crossing/entering data statistics obtained from the Bulgarian Chief Directorate Border Police (Table 1a), in 2017, the Bulgarian-Romanian border was crossed by 3,629,677 vehicles (automobiles and buses, etc.) and 5,113,096 people (pedestrians, traveler(s) in a car and passenger(s) on a train/bus), which is the second highest number of crossings among other borders of Bulgaria. Moreover, when comparing 2017 and 2018, there was an increase in the number of travelers and number of vehicles crossing the border as shown in Table 1b.

Furthermore, as per the updated statistical data received from the Hungarian National Police (HNP), the total number of border crossings in the Romanian section was 23,767,430 people and 10,486,121 vehicles. Moreover, in 2018, the latest Csanádpalota -Nădlac BCP' traffic consisted of 7,215,970 people and 2,727,549 vehicles.

The field visit to the Nădlac and Rusa borders clearly demonstrated the fact that there are long waiting times at the land BCPs. The main reason for this is an increase in the number of travelers and the time taken by border guards to perform the required border checks. Specifically, the main challenges observed during the two visits and based on the interviews that were conducted are:

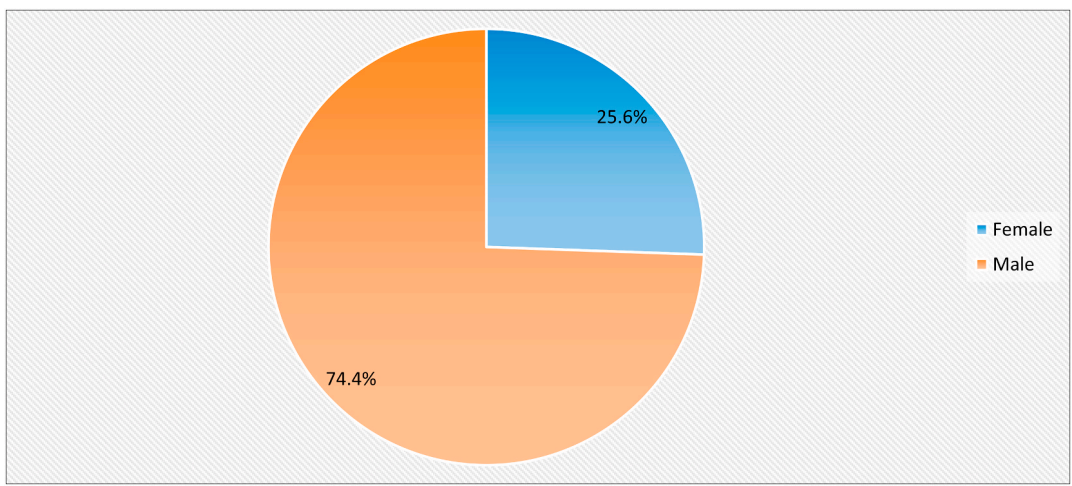

**Fig. 1.** Gender distribution of respondents.

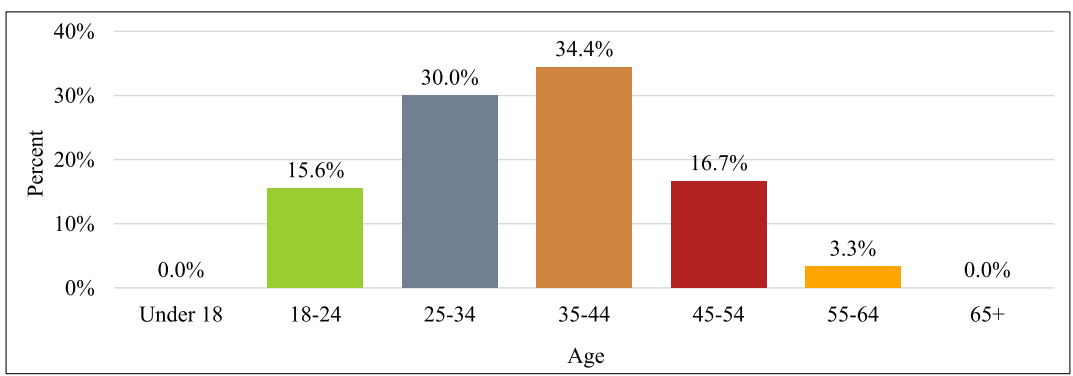

**Fig. 2.** Age distribution histogram of respondents.

**Table 1**
Bulgarian land borders statistical data.

| BORDER | EXIT | ENTER | TOTAL |
|---|---|---|---|
| **(a) Number of crossings between Bulgaria and other countries from January 1, 2017 to December 31, 2017** | | | |
| Bulgaria-Greece | 1311514 | 1984817 | 3296331 |
| Bulgaria-North Macedonia | 1353910 | 1371141 | 2725051 |
| Bulgaria-Serbia | 2330493 | 247143 | 4587636 |
| Bulgaria-Turkey | 3274293 | 3448008 | 6722301 |
| Bulgaria-Romania | 2653559 | 2459537 | 5113096 |
| Danube Bridge-Ruse | 1181767 | 927764 | 2109531 |
| **(b) Number of crossings between Bulgaria and other countries from January 1, 2018 to December 31, 2018** | | | |
| Bulgaria-Greece | 1034235 | 1917485 | 2951720 |
| Bulgaria-North Macedonia | 1405000 | 1421700 | 2826700 |
| Bulgaria-Serbia | 2442856 | 2663335 | 5106191 |
| Bulgaria-Turkey | 4165880 | 4264409 | 8430289 |
| Bulgaria-Romania | 2881383 | 2773380 | 5654763 |
| Danube Bridge-Ruse | 1240722 | 1067309 | 2308031 |

- Long waiting time for vehicles (including cargo vehicles) combined with (occasional) inability of the BCP to serve all vehicles in a timely manner. According to border guards, the number of vehicles crossing borders has tripled since 2010 and the average daily vehicle traffic consists of approximately 8000 cars, 1000 buses and 6000 trucks.
- The use of falsified, damaged and forged documents (e.g., passports, IDs and driving licenses) usually causes delays in BCPs and usually requires more time by border guards to verify documents. According to questions asked during the interview, the average duration for checking a passenger with a car is around two to 3 min, if documents are valid and no alerts are received from the EU and national databases. Meanwhile, the average duration for checking a bus with passengers is about 10–30 min (though in some cases, it can reach 1 h).

Results from the visits and interviews conducted revealed that border guards felt there is an increased need for an improvement in the legacy system which could speed up the checking process and facilitate the overall border-crossing process to reduce waiting times. The improvement should include the integration of mobile devices that can be used for faster and more accurate verification of traveler and car documentation. The border guards felt that the use of mobile devices at BCP booths would relieve the strain on personnel and/or dramatically reduce the time required for vehicular inspection (e.g., engine, interior, and luggage rack checks) and document validation.

Moreover, the researchers observed that there is a need for border guards to use biometrics for traveler identification and verification. Border guards expressed a belief that the integration of biometrics such as fingerprints, facial features, and iris scans would enhance the accuracy of traveler identification and verification, and improve their ability to identify lawful document holders while rejecting impostors according to standard thresholds (e.g., Frontex-European Border and Coast Guard Agency). To the best of the SMILE team's observation and understanding, using biometric technologies would eliminate a considerable amount of stress related to identification and verification procedures faced by travelers and border guards alike. Such technologies would also decrease the amount of tiresome manual border checks required, which would also benefit both groups.

### 4.2. Border guard questionnaires

The results of the questionnaire for border guards (Fig. 3) indicate that apart from IDs (e.g., passports), fingerprint data is the most collected biometric identification at land borders. The results also indicated that the use and enrollment of iris recognition has a very low level of utilization (8.3%). Moreover, the combination of two or more biometric modalities in an identification such as fingerprints and signatures are in use ("Others" in Fig. 3). However, most respondents pointed out that biometric data are collected during second-line checks where a passenger is referred for a more thorough verification carried out in special rooms or offices based on Article 8 of Regulation (EU) 2016/399 (SBC) [9].

Furthermore, the questionnaire results (Fig. 4) showed that cross-match Guardian devices (fixed, not portable devices) are mostly used for control processing at land borders. The survey showed that, currently, land borders do not employ any mobile devices for biometric identification and verification. Moreover, collection and verification of biometrics are done at the BCP booths, and there are no automated gates for biometric data verification at the land border.

Evidence from our observation during the field visit process and interview showed that biometric systems have several advantages to improve border control. However, the main challenge is the duration of the process of collecting and verifying the data (in some cases as described by the interviewee). According to the questionnaire results, the time spent on each traveler (pedestrian) is anywhere from 1 to 12 min if data is valid and no alerts are received from the EU and national databases. Similarly, waiting time per vehicle is 3–15 min in cases of personal cars where no suspicion is raised and 3 min to hours (where the number of hours was not clearly indicated) in cases of passenger buses and trucks. According to the SMILE interview, the waiting time for trucks might be up to 2 h. Considering the high number of travelers as shown in Section 4.1 and assuming the average waiting time, we could say that travelers will be facing delays, and that long of cars, buses and trucks would be formed. Thus, as we mentioned earlier, border officers might not be able to serve all travelers/vehicles in sufficient time.

This increase of waiting time happens due to the requirements and process of border checks. According to Regulation (EU) 2016/399 (SBC), border checks refer to the checks carried out at border crossing points, to ensure that persons, including their means of transport and the objects in their possession, may be authorized to enter the territory of the Member States or to leave it. Based on the SMILE survey result, 33% of respondents (border guards) indicated that drivers and passengers are asked to leave the car during border checks (e.g., drivers might be asked to show the car's trunk). Also, in some cases, passengers in buses might be asked to get out of the bus for individual checks. Whereas 83% of respondents indicated that there are no control differences for passengers in different vehicle types.

Another challenge results from situations where it is difficult to collect biometrics from passengers e.g., biometrics from children and people with disabilities. Questionnaire results illustrated that currently there are no specific procedures for people traveling with children or as a family. The only specific measures for families mentioned by some of the respondents is that "If a child (or children) is accompanied by just one parent, that parent is required to show a document proving that the other parent consents to the child leaving the country."

Additionally, challenges occur due to false reporting of identities at the land border. Questionnaire results illustrated that there is a high number of false reporting of identities and forged traveler/vehicular documents at the land border. However, the approximate ratio of false reporting of identities/documents to the actual reporting of identities was not clearly identified due to the small sample size. The result of the questionnaire also indicated that (as mentioned in Section 1) biometric information systems such as SIS and VIS play an increasingly important role in the establishment and verification of traveler identity. The main objectives of such information systems are to share and transfer all required personal data among BCPs to be able to perform background checks on individuals.

Evidence from the survey shows that biometric technologies have several advantages to improve border control. The integration of biometrics at border control provides benefits for travelers, political entities (states), authorities responsible for border control and individual border guards. The most important benefits identified by the border guards were accuracy, integrity, robustness, and efficiency. Accuracy of a traveler's identification means the ability to recognize genuine documents and detect counterfeits. Biometric systems can help to maintain integrity by verifying that collected data (printed and digital) has been produced by an authorized organization (as opposed to a fictitious one, for example) and that it has not been subject to any unauthorized alteration (e.g., extension of passport validity or introduction of a visa sticker or entry/exit stamp that has not been done by an issuing state, consulate, or border authority). Robustness of biometric systems is related to the ease with which they can be operated, maintained, updated, replaced, redeployed, or decommissioned when compared to border control units/booths that rely on human agents. The efficiency of biometric systems is related to the processing capacity of ABC and its sustainability over time, because unlike human border guards, ABCs do not experience fatigue.

### 4.3. Traveler questionnaires

A majority of the respondents (70.4%) declared that they had experienced long queues and wait times at border crossings (Fig. 5). Whereas, 29.6% reported that they have not experienced this problem. Long queue and waiting in border crossing are the main concern for travelers. Among the 57 respondents that reported the long queues, 86% declared that this was primarily a result of the high flow of people through the border (Fig. 6), while 43.9% stated that it was a result of passport processing and document control at the BCPs. Furthermore, 7% of respondents felt that long waiting times were due to a lack of documents/procedure on their own behalf, and 21% felt they were due to system failure or a lack of organization by BCPs. Related to the major problems (Fig. 7) faced at the BCPs, communication issues with border guards and not having access to enough information was reported frequently (43.1% each). About 16.9% of respondents reported reliability and fair treatment as issues, while 26.2% of them stated that they did not understand the information related to border procedures.

When it come to the use of biometrics for border control, as shown in Fig. 8, 46.7% reported that they worried about personal data theft or/and misuse due to unauthorized disclosure of sensitive information. Also, 46.7% felt that the extensive data collection could lead to privacy violation, whilst 30.7% felt uncomfortable with the police recording and processing their physical features.

Based on these statistics, we can conclude that the use of biometric identifiers was a source of concern to respondents in terms of their privacy and personal data security. Interestingly, 80.9% of the respondents agreed that the border crossing checks and flow had to be improved using automating border control checks, while 10.1% were neutral and only 8.9% disagreed (Fig. 9). Based on these figures, we can conclude that there is a high demand from travelers to improve crossing speed and flow.

However, 65.2% of the respondents envisaged difficulties when using automatic border control systems. On the other hand, 34.8% did not think that they would have any difficulties in case an automatic border control system was implemented. Among the 65.2% of respondents who reported problems in using automatic border control systems, 62.3% reported lack of awareness and knowledge about technologies as the problem (Fig. 10). Also, travelers were concerned about security vulnerabilities and privacy controls (59%). Problems related to the complexity of technology were only reported by 44.3% of respondents, while difficulties due to their own ability to maintain the technology with respect to fixing, updating, extending, operating and servicing it was reported by 42.6%.

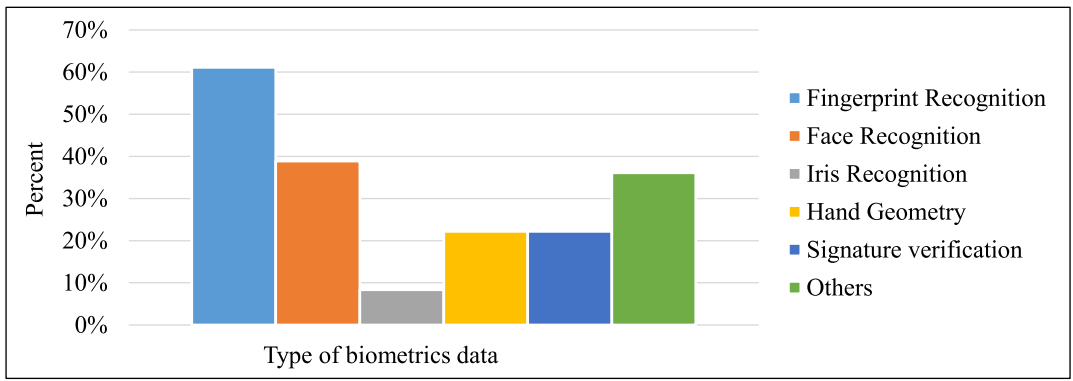

**Fig. 3.** Type of biometric data currently being collected at land borders.

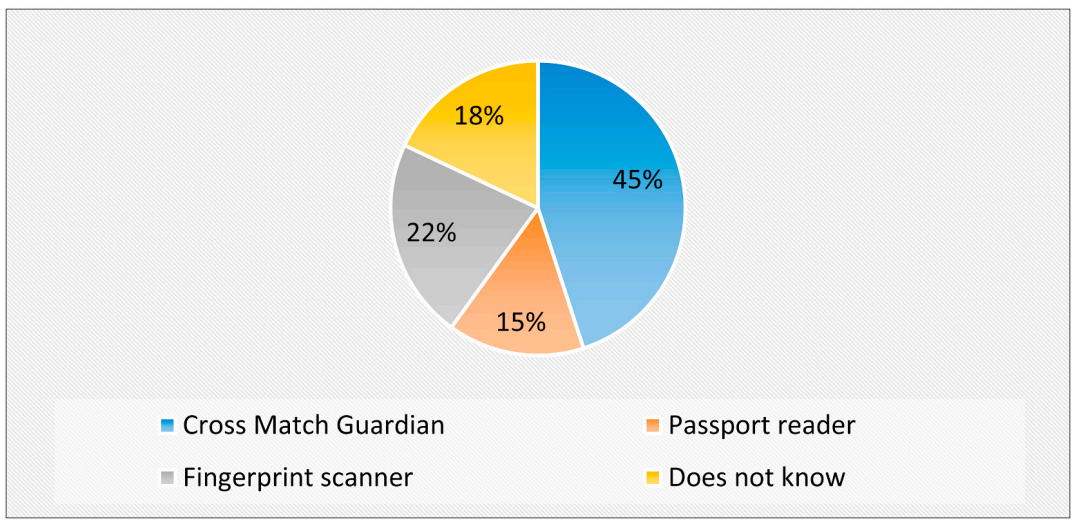

**Fig. 4.** Type of device used for biometric data collection.

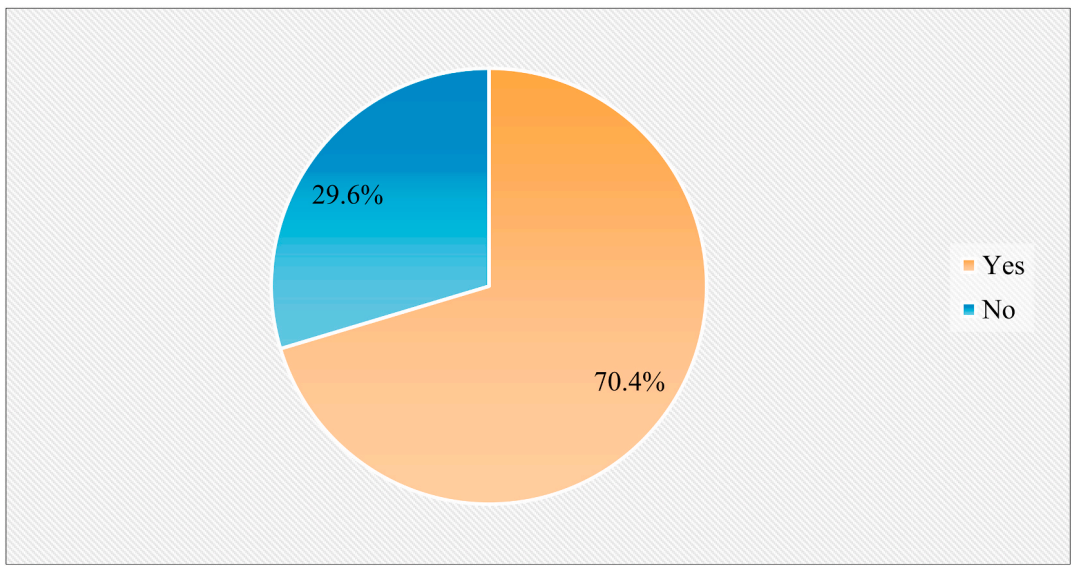

**Fig. 5.** Experience of long queues and waiting times at land borders.

The main benefits that travelers were expecting from an automated border control system were speeding up the border control processes (71.4%) and reduced waiting time at land border (66.7%) as shown in

Fig. 11. Travelers expected less improvement in preventing misuse of personal information and better privacy and security control of personal data by using automated border control system.

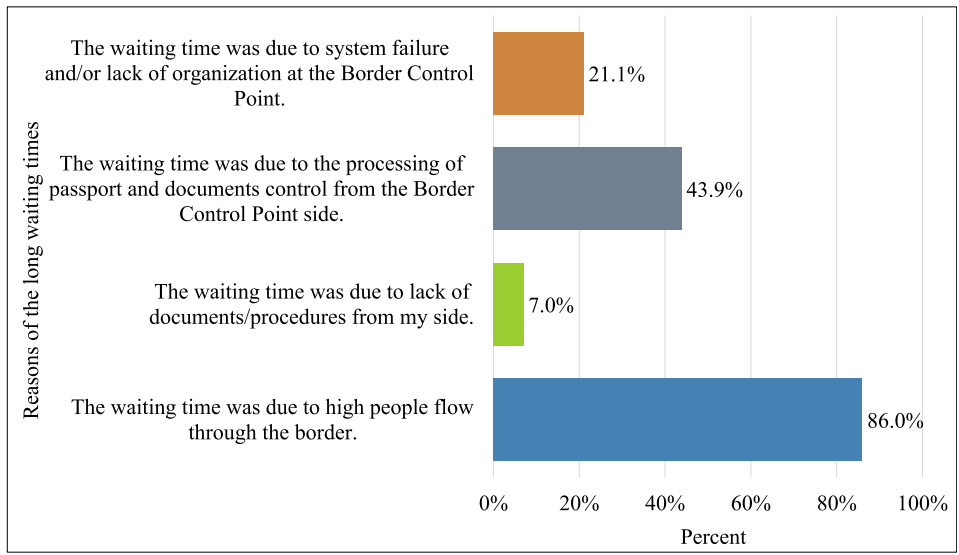

**Fig. 6.** Reasons for long waiting times.

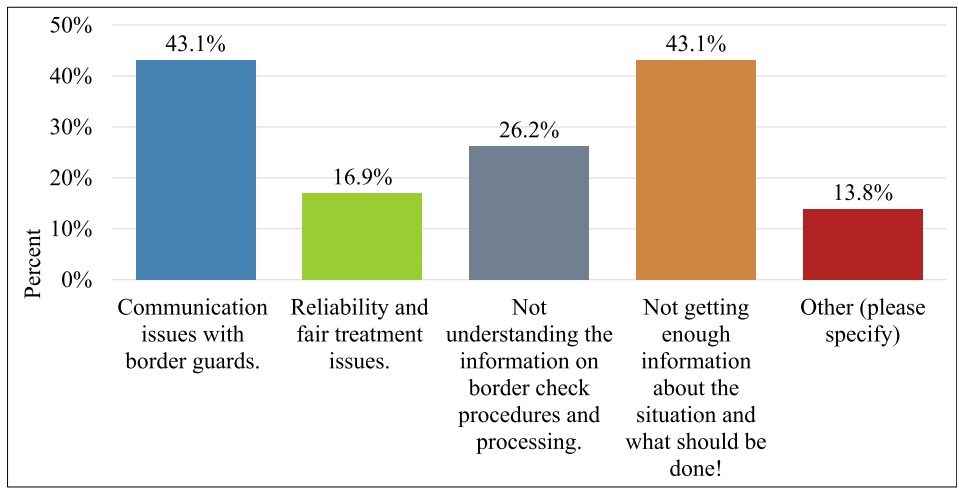

**Fig. 7.** Major problems at border crossing points.

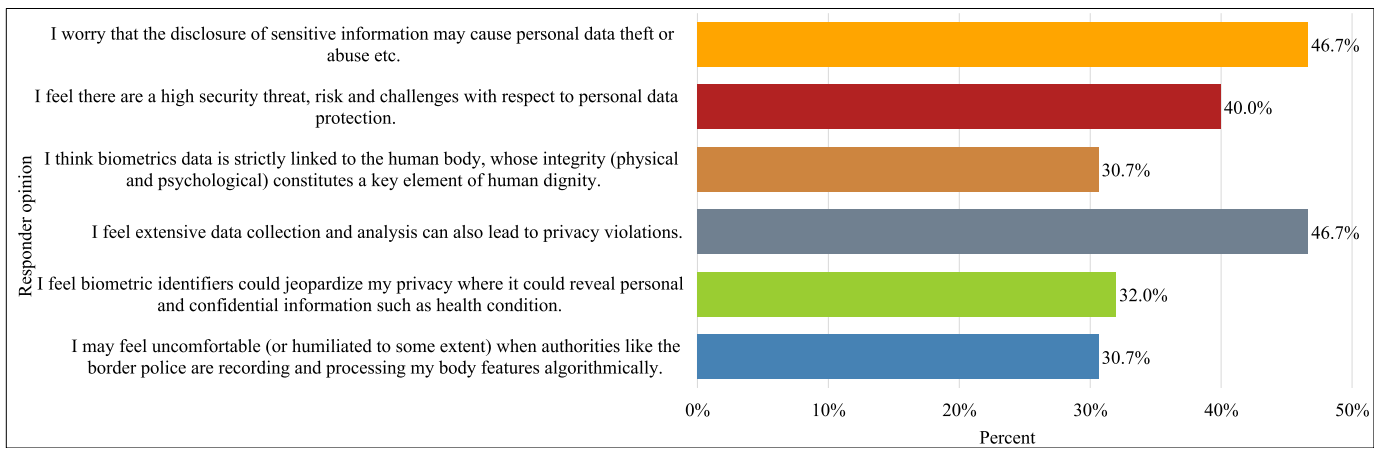

**Fig. 8.** How the use of biometric identifiers could jeopardize privacy and personal data security.

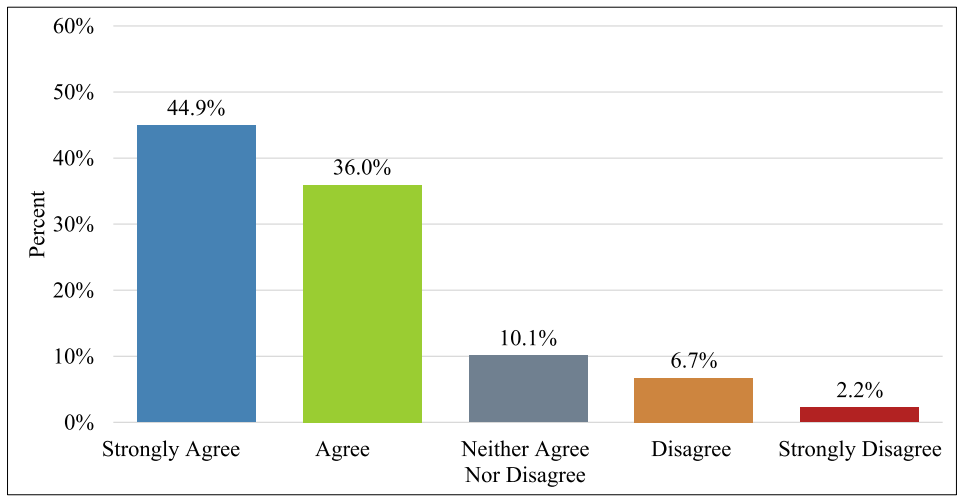

**Fig. 9.** Responses to whether there is a need for improving crossing speed and flow by automating border control checks processing.

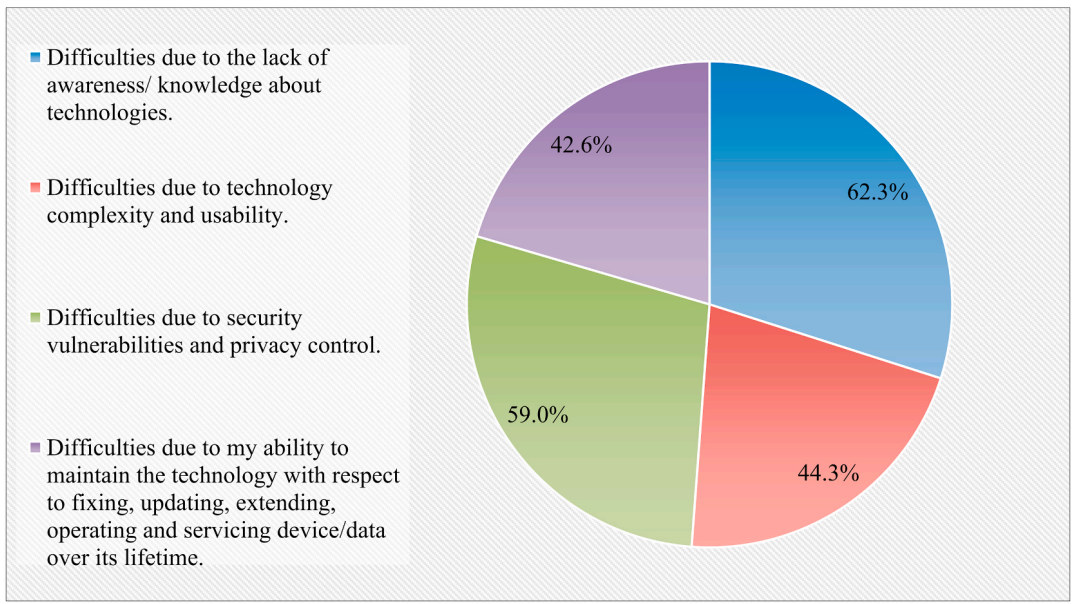

**Fig. 10.** The respondents' difficulties associated with using an automated border control system.

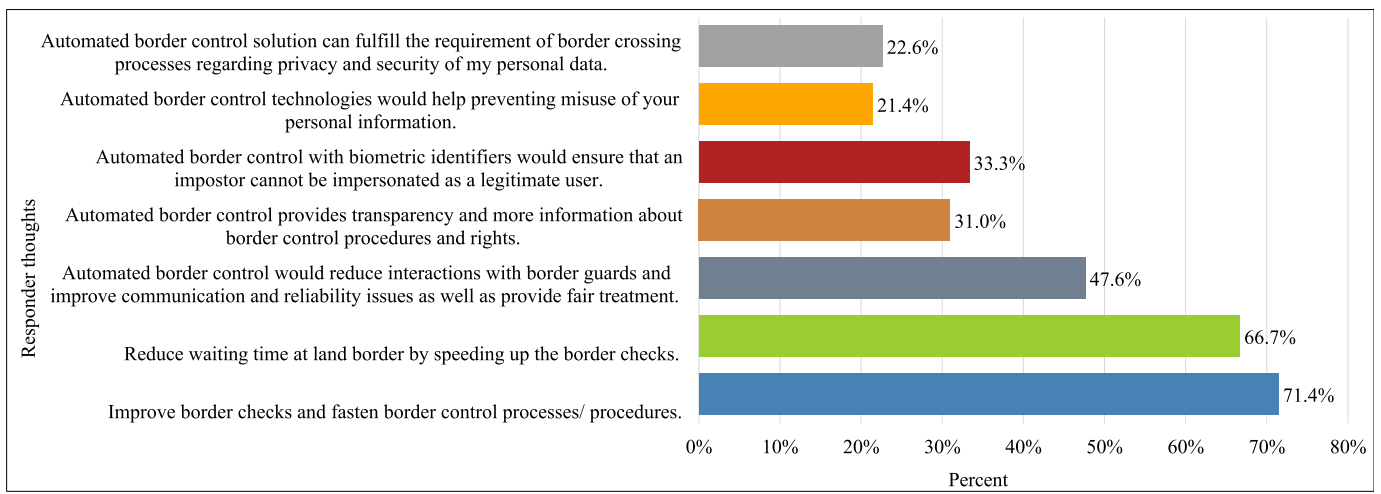

**Fig. 11.** Thoughts about how biometric identifiers and mobile technologies would improve border control.

## 5. Discussion and recommendation

### 5.1. Border authorities' views

The growing flow of land travelers along with rising global mobility are significantly increasing the processing time to cross borders. It is understood that the current land border control infrastructures are basically relying on stand-alone (fixed) devices where most border checks such as identity control and flow management are carried out in BCP booths. Our survey showed that these infrastructures (while having worked well for several decades), are now sources of many new challenges, and border authorities are being confronted with difficulties in managing the flow of travelers at borders and in fulfilling the obligations set by the Schengen Borders Code (SBC). In order to streamline operations at borders, EU and border authorities are calling for continuous upgrades (or even replacement) of the existing identification, verification and cross-border management systems to facilitate and speed up border crossings in a secure manner.

In this context, border authorities anticipate that seamless integration of biometrics-based mobile devices with existing border control infrastructures may pave the way towards faster border crossing, enhanced security levels, better control and management and much lower costs. The main purpose of biometrics technologies (as claimed by SMILE end users) is to improve the effectiveness and efficiency of border controls, provide more accurate data on entry and exit, speed up border crossings, and ultimately to lead to seamless crossing of borders and security checks for the vast majority of travelers. Moreover, biometrics technologies would help identify travelers entering/leaving the country in a more precise and reliable way.

### 5.2. Travelers' views

Many travelers have already been exposed to biometrics technologies in the context of air travel and immigration (e-passport). The main benefits that travelers are expecting from the use of biometrics technologies include speeding up the border control processes and reducing their waiting time at the border. Biometrics technologies will also facilitate easy and fast border crossing for travelers without the need to wait long in queues. Moreover, the ability for these technologies to fight terrorism and serious crime and ensure a high level of internal security will combat irregular migration and raise border management security levels. These will have indirect but arguably very positive effects on public citizens.

On the other hand, the deployment of biometrics technologies is acknowledged to potentially raise critical ethical and social concerns (e. g., concerns regarding the fundamental rights of privacy and data protection). The combination of biometric information systems and mobile technologies particularly lead to increased surveillance including collecting and storing facial images, license plates, fingerprints, etc. as individuals cross borders, apply for visas or request asylum. Also, public concerns about the lack of transparency and lawfulness of data processing could lead to risks to the rights of a person such as discrimination against individuals, social exclusion and intrusion into an individuals private life. These can influence the social acceptability of biometric identification and verification methods.

Furthermore, it is relevant to recognize that in the present debate, the vast majority of travelers have no experience using biometric technologies (Fig. 10). Therefore it remains unclear what these people truly understand about the proposed changes. Thus, performance expectations from travelers perspectives need to be explored in detail. Furthermore, border authorities must ensure the credibility of all these technologies before deploying them as part of border management systems. There are no commonly accepted, standardized proportionality tests related to the use of biometrics in the context of border security, however there are supporting analogies in guidelines and recommendations as well as multiple EU research projects (e.g., PERSONA[2]) on the way to elaborate a solution.

### 5.3. Recommendation

Biometric data is considered as a special category of personal data (Article 9 of GDPR) and its processing is prohibited (Article 9(1) of GDPR [11,27]) unless processed in accordance with Article 9(1) and (2) of GDPR. For example, processing must be done with explicit consent of the data subject, whose vital interests must be protected. Meanwhile, border authorities carry out data processing based on a partially different legal framework, where many prohibitions and safeguards of GDPR are derogated by other EU regulations or by national implementation of EU directives (e.g. no informed consent required, transparency is not always obligatory), however this does not lift the requirement of proportionality and professionalism, a burden that is borne by the data controller authority.

GDPR and other frameworks for data protection such as the Bulgarian Personal Data Protection Act encourage the development of biometric technologies (e.g., SMILE) practices that are premised on the understanding that data privacy and protection are not just a matter of simple compliance, but that they are a necessity that all policymakers and actors must take into consideration. Therefore, policy for biometric data collection and processing must be based on legal principles and involve the participation of diverse actors. Also, the deployment of biometrics technologies must involve control mechanisms that help ensure transparency and accountability.

It is concluded that the implementation of biometrics technologies at borders must be in compliance with fundamental rights, in particular in relation to privacy and data protection. Biometric technologies platforms must be designed to support privacy and data protection compliant biometrics systems. In order to ensure respect for data protection and privacy, many factors must be considered, as follows:

- **Data confidentiality**: Data must be protected both logically (such as preventing data from unauthorized access) and physically (such as data loss, like the irreversible failure of a storage device). A data breach may affect data that is business-sensitive in nature, and the loss of this data may lead to one or more types of losses, for example, loss of competitive advantage and non-compliance fines. Many mechanisms and techniques can be used to ensure data confidentiality including data encryption, data minimization, data anonymization, and data pseudonymization. Requirements to safeguard confidentiality are as follows:
  - Prevention of unauthorized access to data.
  - Restriction of access for authorized personnel according to their professional needs.
  - Maintenance of an overview (logs) of everyone who has gained access to data.
- **Data integrity**: Refers to maintaining the accuracy and validity of data throughout its life cycle, ensuring that it is not altered or destroyed in an unauthorized manner. Data integrity assurance must be enforced via cryptographic controls for detection of integrity violations. Requirements to safeguard integrity are as follows:
  - Prevention of accidental or unauthorized alteration or erasure.
  - Ensuring that data are accurate and where necessary kept up to date.
  - Prevention of copies of data from becoming a source of outdated information.
- **Data availability**: Maintaining data availability is essential for the performance and business continuity of border authorities. If border authorities were to lose access to border control data, IT operations such as SMILE could cause a halt, resulting in significant or even irreversible consequences which they would only be able to overcome with serious difficulties if at all. Requirements to safeguard availability are as follows:

- Ensuring that data are available according to professional need.
- Ensuring that appropriate technical and organizational measures are in place.
- Enabling prevention, detection, scalability, management and restoration.
- Ensuring that information systems are available in accordance with the organization's availability requirements.

---

[2] http://persona-project.eu/.

Moreover, to avoid the negative consequences of the technology, we recommend that the deployment of biometric technology platforms consider travelers with special needs/categories including (to name a few):

- Travelers with temporary injuries who might have difficulties to provide biometric samples due to a temporary wound.
- Travelers with total permanent disability who have difficulty freely moving their limbs due to sensory damage and/or muscle damage (e. g., during fingerprint verification, travelers with a hand disability may lack the ability to place the required finger where necessary and keep it steady for a sufficient time on the fingerprint scanner).
- Travelers with technological illiteracy who lack knowledge of technology/tools (e.g., elderly people) and have difficulties using and interacting with biometric systems such as e-gates.

### 5.4. Limitation of the study

The main limitation of this study is the small sample size used for the analyses. The number of responses received were insufficient to form a conclusive opinion about the primary stakeholders' views on the use of biometrics technologies in border control. Therefore, to generalize the results for a larger group, the study needs a larger sample size. Moreover, future work may involve performing an assessment based on privacy and data protection principles and regulations (e.g., GDPR and Directive 2016/680). This assessment would ensure the minimization of access to data based on the need to know principle, the proportionality between the amount of information collected and retained and the objective of the technology, the protection of the data collected and the minimization of negative outcomes in the event of a data breach.

### 6. Conclusion

The aim of this study was to build a general understanding of the need to deploy biometric technologies at land borders and how relevant legal and policy frameworks such as GDPR might affect the development of biometric technologies as a border control system. It is understood that biometric technologies are at the heart of current development of border checks. The European States are initiating an increasing number of projects that make use of biometric technologies. More significantly, in many settings across Europe potential or actual uses include traveler identification and verification by storing data in EU biometric information systems, such as VIS, SIS II and the Entry-Exit System (EES). Borders authorities utilize biometric technologies at border control to achieve an automated, rapid and highly secure self-service clearance process, such that increasing passenger throughput does not compromise border control reliability. Moreover, the technology provides computerized decision-making support to border control authorities and increases the reliability and efficiency of border control measures. However, the efficacy of biometric technologies can be affected by the ethical, cultural, social, and legal considerations that shape how people engage and interact with biometric systems. On one hand, biometric data can be used to recognize individuals automatically with greater accuracy. On the other hand, a misuse of such biometric data can have dangerous consequences which pose several security and privacy challenges such data destruction and/or unauthorized disclosure of or access to personal data.

### Acknowledgement

This work is carried out in the EU-funded project SMILE (Project ID: 740931), [H2020-DS-2016-2017] SEC-14-BES-2016 towards reducing the cost of technologies in land border security applications.